%% file: sbmglm.tex
\documentclass[imslayout,preprint]{imsart}
\usepackage[left=1.25in,right=1.25in,top=1in,bottom=1in]{geometry}

\usepackage[OT1]{fontenc}
\usepackage{ae,aecompl}
\usepackage{graphicx}
\usepackage{amsthm,amssymb,amsmath,amsfonts}
\usepackage{enumerate}
\usepackage{algorithmic}
\usepackage{algorithm}
\usepackage{natbib}
\usepackage{hyperref}
\usepackage{caption}
\usepackage{subcaption}

\startlocaldefs%
\newtheorem{theorem}{Theorem}
\newtheorem*{theorem*}{Theorem}

\theoremstyle{definition}

\theoremstyle{remark}

\def\argmin{\mathop{\rm arg\,min}\limits}%
\def\argmax{\mathop{\rm arg\,max}\limits}%
\def\iid{\ensuremath{\stackrel{\text{\tiny iid}}{\sim}}}%
\def\ind{\ensuremath{\stackrel{\text{\tiny ind}}{\sim}}}%
\let\hat\widehat%
\let\tilde\widetilde%
\def\given{{\,|\,}}
\def\Pr{{\ensuremath{\mathbb P}}}%
\def\Exp{{\ensuremath{\mathbb E}}}%
\endlocaldefs%
\begin{document}

\begin{frontmatter}

\title{Bayesian Degree-Corrected Stochastic Blockmodels for Community
Detection}
\runtitle{Bayesian Deg.-Corr. SBM for Comm. Detection}

\author{\fnms{Lijun} \snm{Peng}\corref{}%
\ead[label=e1]{ljpeng@math.bu.edu}}
\and
\author{\fnms{Luis} \snm{Carvalho}%
\thanksref{t2}\ead[label=e2]{lecarval@math.bu.edu}}
\thankstext{t2}{Supported by NSF grant DMS-1107067.}
\runauthor{L. Peng and L. Carvalho}
\affiliation{Boston University}
\address{Department of Mathematics and Statistics\\
Boston University\\
111 Cummington Mall\\
Boston, Massachusetts 02215\\
\printead{e1,e2}}

\begin{abstract}
Community detection in networks has drawn much attention in diverse fields,
especially social sciences. Given its significance, there has been a large
body of literature with approaches from many fields. Here we present a
statistical framework that is representative, extensible, and that yields an
estimator with good properties. Our proposed approach considers a stochastic
blockmodel based on a logistic regression formulation with node correction
terms. We follow a Bayesian approach that explicitly captures the community
behavior via prior specification. We further adopt a data augmentation
strategy with latent P\'olya-Gamma variables to obtain posterior samples. We
conduct inference based on a principled, canonically mapped centroid estimator
that formally addresses label non-identifiability and captures representative
community assignments. We demonstrate the proposed model and estimation on
real-world as well as simulated benchmark networks and show that the proposed
model and estimator are more flexible, representative, and yield smaller error
rates when compared to the MAP estimator from classical degree-corrected
stochastic blockmodels.
\end{abstract}


\end{frontmatter}

\input{intro-carvalho.tex}

\input{bsbmodel-carvalho}

\input{postsamp}

\input{postinf}

\input{experiment-carvalho}
\input{discussion-carvalho}

\input{appendix-carvalho}

\bibliographystyle{chicago}
\bibliography{sbmglm}

\end{document}

%% file: intro-carvalho.tex
\section{Introduction}
Networks can be used to describe interactions among objects in diverse fields
such as physics~\citep{newman06}, biology~\citep{hancock10}, and especially
social sciences~\citep{zachary77,adamic05}. In network theory, objects are
represented by \emph{nodes} and their interactions by \emph{edges}. Clusters
of nodes that share many edges between them but that, in contrast, do not
interact often with nodes in other clusters can be thought of as
\emph{communities}.
This characterization follows a traditional approach in social sciences that
aims at discerning the structure of a network according to relationship
patterns among ``actors'', e.g.\ friendship or collaboration. These interaction
patterns may reflect ``assortativity'', a concept that originated in the
ecological and epidemiological literature~\citep{albert02}: it refers to the
tendency of nodes to associate with other similar nodes in a network. Among
measures of similarity, the degree of a node is of common interest in the
study of assortativity in networks~\citep{newman02A, newman03A, vazquez03},
that is, assortative networks usually show a preference for high-degree nodes
to connect to other high-degree nodes. We expect in some applications that
actors exercise assortativity and prefer to group themselves according to
similarity or kinship in communities, and so communities are \emph{dense} in
within-group associations but \emph{sparse} in between-group interactions.
Thus, not surprisingly, community detection has sparked great interest in many
fields where recent applications aim at characterizing the structure of a
network by detecting its communities.

There have been many approaches to address community detection (see
Section~\ref{sec:priorwork} for a more thorough review), but a common modeling
choice is to treat actors as behaving similarly given their respective
communities. This structural equivalence assumption is at the core of
\emph{blockmodels}~\citep{lorrainwhite71}, which were later extended to
stochastic blockmodels~\citep{holland81,fienberg85}.
Here, to tackle community detection, we adopt a hierarchical Bayesian
stochastic blockmodel where group labels are random.
We contend that a suitable prior specification is essential to accurately
characterize assortative behavior, and thus that a Bayesian approach is
essential to community detection (see, e.g., the examples in
Section~\ref{ssec:examples}.)
Our results can be connected to the work of \citet{nowicki01},
\citet{karrernewman11} and \citet{hofman08} but we make two important
distinctions: (i) we capture community behavior by explicitly requiring that
the probability of within-group associations is higher than between-group
relations; and (ii) we address parameter and label non-identifiability issues
directly by remapping configurations to a unique canonical space. The first
point is important in light of the examples in the last section. The second
point allows us to sample from the posterior space of label configurations
more efficiently and to formally define an estimator based on a meaningful
loss function. Moreover, our model can be related to the work of
\citet{mariadassou10} and \citet{vu13} as they are all based on
exponential-family clustering frameworks, but our model is different from
theirs in two respects besides the two points just mentioned: (i) we make
exact inference by adopting latent variables, rather than adopting approximate
variational approaches; and (ii) we add more flexibility by requiring
hyper-prior structure on model parameters controlling degree correction. 

More specifically, we make the following contributions:
\begin{enumerate}[(1)]
\item We propose a Bayesian degree-corrected stochastic blockmodel for
community detection that explicitly characterizes community behavior. We
discuss this new model and how we account for parameter non-identifiability in
Section~\ref{sec:bsbm}.

\item We treat label non-identifiability issues by defining a canonical
projection of the space of label configurations in Section~\ref{sec:label}.

\item We develop an efficient posterior sampler by identifying good initial
configurations through approximate mode finding and then exploring a Gibbs
sampler based on a data augmentation strategy in Section~\ref{sec:postsamp}.

\item We propose a \emph{remapped} centroid estimator for community inference
in Section~\ref{sec:postinf}. This new estimator is based on Hamming loss and
is arguably a good representative of a projected space of label configurations.
\end{enumerate}

In Section~\ref{sec:exp} we show that our proposed method is efficient and
able to fit medium-sized networks with thousands of nodes in reasonable time.
Moreover, we show that our proposed estimator yields, in practice, smaller
misclassification rates due to a more refined loss function when compared to
the ML-based estimators. Finally, in Section~\ref{sec:disc}, we offer some
concluding remarks and directions for future work.

\section{Prior and Related Work}
\label{sec:priorwork}
There is a large body of literature in community detection, given its
significance and interest. Traditional methods include graph partitioning
\citep{Kernighan70,Barnes82}, hierarchical clustering \citep{hastie01}, and
spectral clustering \citep{Donath73,von07,rohe11};
while these methods are heuristic and thus suitable for large networks, they
do not address directly community detection but aim instead at partitioning
the network according to edge densities between groups and thus identifying
connection ``bottlenecks''.

The concept of \emph{modularity} better captures community structure by also
taking within-group edge densities into account \citep{newman04a,newman06}.
Optimization methods based on modularity can then be used to detect
communities, but since modularity optimization is
NP-complete \citep{brandes07}, interest lies mostly in approximated methods
such as the greedy method of \citet{newman04b} and extremal
optimization \citep{duch05, bickel09}. However, there are still drawbacks:
methods based on modularity may fail in detecting small communities and thus
exhibit a ``resolution limit'' \citep{fortunato07}.
Latent space network models~\citep{hoff02}, latent variable
models~\citep{hoff05}, and latent position cluster models~\citep{hancock07}
assume that the probability of an interaction depends on node-specific latent
factors such as the distance between two nodes in an unobserved continuous
``social space''; these models are generalizations of exponential random graph
models [ERGMs; see~\citep{robins07}] where community structure is assumed from
cluster structure in the latent space.

There are many other methods to mention [see, for example, the review
in \citep{parthasarathy11}], but we focus on parametric statistical approaches
where inference on community structure is based on an assumed model of
association. The motivation is that since there are many possible community
configurations, that is, assignment of actors to communities, we want to not
only infer communities, but to also assess how likely each configuration is
according to the model.

The first endeavors in such parametric models---albeit not in community
detection---are the $p_1$ exponential family models due to~\citet{holland81}.
These models follow a log-linear formulation~\citep{fienberg81} with
parameters that are related to in- and out-degrees and edge densities. Later, these models were extended to incorporate actor and group parameters~\citep{fienberg85,tallberg05, daudin08}. \citet{wang87} further adapted the models to consider a block
structure through \emph{stochastic blockmodels} [SBMs~\citep{holland83,anderson92}], yielding $p_1$ blockmodels. \citet{zanghi10}, \citet{mariadassou10} and \citet{vu13} proposed scalable approximate variational approaches based on modified version of those $p_1$ (block)models. 

Stochastic blockmodels explore a simpler model structure where the probability
of an association between two actors depends on the groups to which they
belong, that is, two actors within the same group are stochastically
equivalent.
\citet{karrernewman11} developed an SBM that allows for
\emph{degree-correction}, that is, models where the degree distribution of
nodes within each group can be heterogeneous. \citet{celisse12}, \citet{choi12} and \citet{bickel13} addressed the asymptotic inference in SBM by use of maximum likelihood and variational approaches. More flexible approaches generalize the SBM by adopting a hierarchical
Bayesian setup that regards probabilities of association as random and group
membership as latent variables~\citep{snijders97,nowicki01,hofman08}.
As in all latent mixture models, label non-identifiability is a known
problem since multiple label assignments yield the same partition into
communities; ultimately, we only care if two actors are in the same community
or in different communities. It is also possible to incorporate node
attributes in the model~\citep{kim11,hoff13} and to allow actors to belong to
more than one community~\citep{airoldi08}.

%% file: bsbmodel-carvalho.tex
\section{A Bayesian Stochastic Blockmodel for Community Detection}
\label{sec:bsbm}

Under our community detection setup we assume a \emph{fixed} number of groups
$K \ge 2$ and we are given, as data, a matrix $[A]_{ij}$ representing
relationships between ``actors'' $i$ and $j$ in a network with $n > K$ nodes.
We represent the assignment of actors to communities through
$\sigma : \{1,\ldots,n\} \mapsto \{1,\ldots,K\}$, a vector of
\emph{labels}: $\sigma_i = k$ codes for the $i$-th individual belonging to the
$k$-th community.

A simple stochastic blockmodel specifies that the probability of an edge
between actors $i$ and $j$ depends only on their labels $\sigma_i$ and
$\sigma_j$, and that $\sigma$ follows a product multinomial distribution:
\begin{equation}
\label{eq:simplesbm}
  \begin{aligned}
    A_{ij} \given \mathbf{\sigma}, \theta \ind
    \mbox{\sf Bern}\big( \theta_{\sigma_i \sigma_j} \big),
    &\qquad i,j = 1, \ldots, n, i < j, \\
    \sigma_i \iid \mbox{\sf MN}(1; \mathbf{\pi}),
    &\qquad i = 1, \ldots, n,
  \end{aligned}
\end{equation}
where $\mathbf{\pi}$ is a vector of prior probabilities over
$K$ labels, parameter $\theta_{kk}$ is the ``within'' probability of a
relationship in community $k$, and $\theta_{kl}$ is the ``between''
probability of a relationship for communities $k$ and $l$, $k, l = 1, \ldots,
K$, $k < l$. If we define $\theta_w \doteq \theta_{11} = \cdots = \theta_{KK}$
and $\theta_b \doteq \theta_{12} = \cdots = \theta_{K-1,K}$, we have a simpler
model with single within and between probabilities~\citep{hofman08}.

We regard SBMs as log-linear models and exploit this formulation to
define a \emph{node-corrected} SBM by
\begin{equation}
\label{eq:logitsbm}
A_{ij} \given \sigma, \gamma, \eta \ind
\mbox{\sf Bern}\big( \text{logit}^{-1}(\gamma_{\sigma_i\sigma_j}
+ \eta_i + \eta_j) \big)
\end{equation}
where, in logit scale, parameters $\gamma$ capture within and between
community probabilities of association and node intercepts
$\eta = (\eta_1, \dots, \eta_n)$ capture the expected
degrees of the nodes. To avoid redundancies, we only code $\gamma_{kl}$ for $k
\le l$. We note that without $\eta$, model~\eqref{eq:logitsbm} is equivalent
to model~\eqref{eq:simplesbm} with $\gamma_{kl} = $logit$(\theta_{kl})$. We
also remark that we call the above model node-corrected, which is arguably
more suitable for a broader generalized linear model formulation; in Karrer
and Newman's approach the observed $A_{ij}$ follow a Poisson distribution, and
so $\eta$ is related to expected log degrees, and hence their
degree-correction denomination~\citep{karrernewman11}.

\subsection{Parameter Identifiability}
In what follows, to simplify the notation we group $\beta = (\gamma, \eta)$
and define the design matrix $X$ associated to model~\eqref{eq:logitsbm}
such that
\[
A_{ij} \given \sigma, \beta \iid \text{\sf Bern}\big(
\text{logit}^{-1}(x_{ij}(\sigma)^\top \beta)\big).
\]
Note that we make explicit the dependence of each row $x_{ij}$ on the labels
$\sigma$. Model~\eqref{eq:logitsbm} has then $\binom{K}{2} + K + n$
parameters, but the next result shows that only $\binom{K}{2} + n$ parameters
are needed for the model to be identifiable if each community has at least two
nodes (the proof is in Appendix~\ref{sec:proofdesign}.)

\begin{theorem}
\label{thm:design}
The design matrix $X$ associated with model~\eqref{eq:logitsbm} has the
following properties:
\begin{enumerate}[(1)]
\item It has $K$ linearly dependent columns.
\item It is full column-ranked if and only if each community has at least two
nodes.
\end{enumerate}
\end{theorem}

Based on these two criteria, to attain an identifiable model we remove $K$
parameters from $\gamma$ and modify the prior on $\sigma$ to a constrained
multinomial distribution,
\[
\Pr(\sigma) \propto \prod_{k=1}^K I(N_k > 1)
\prod_{i=1}^n \pi_k^{I(\sigma_i = k)},
\]
where $I(\cdot)$ is the indicator function and $N_k = \sum_i I(\sigma_i = k)$
is the number of nodes in community $k$. There are still problems with label
identifiability that we address by label remapping in the
Section~\ref{sec:label}; for now, to allow for a straightforward remapping of
community labels, we just set
\begin{equation}
\label{eq:gammaconst}
\gamma_{11} = \cdots = \gamma_{KK} = 0
\end{equation}
to remove the redundant $\gamma$ parameters.

\subsection{Hierarchical model for community detection}\label{sec:hiemodel}
We attain a more realistic model by further setting a hyper-prior
distribution on
$\gamma = (\gamma_{12}, \ldots,$ $\gamma_{K-1, K})$, $\eta$, and
$\pi$,
\begin{equation}
\label{eq:prior}
  \begin{aligned}
  \beta = (\gamma, \eta) &\sim
    I(\gamma \leq 0)\cdot N\Big(0, \tau^2 I_{n+\binom{K}{2}}\Big), \\
  \pi &\sim \mbox{\sf Dir}(\alpha_1, \dots,\alpha_K),
  \end{aligned}
\end{equation}
where $\tau^2$ controls how informative the prior is. The prior on $\gamma$
and $\eta$ can be seen as a ridge regularization for the logistic regression
in~\eqref{eq:logitsbm}.
The constraint $\gamma \leq 0$ in this stochastic blockmodel is essential to
community detection since we should expect as many as or fewer edges between
communities than within communities on average, and thus that the log-odds of
between and within probabilities is non-positive. The conjugate prior on $\pi$
adds more flexibility to the model, and is important when identifying
communities of varied sizes and alleviating resolution limit issues.

\section{Label Identifiability}
\label{sec:label}
Since the likelihood in~\eqref{eq:logitsbm} only considers if individuals are
in the same community or not, labels are not identifiable due to this
stochastic equivalence. Moreover, if $\pi$ follows a strongly
informative symmetric Dirichlet, $\alpha = W \cdot1_{K}$ with $W$ large, then
the marginal prior on $\sigma$ is approximately non-identifiable:
\[
\Pr(\sigma) = \int \Pr(\sigma \given \pi) \Pr(\pi) \text{d}\pi
= \frac{\prod_k \Gamma(N_k + W) / \Gamma(W)}%
{\Gamma(n + KW)/\Gamma(KW)}
\approx \frac{\prod_k W^{N_k}}{(KW)^n} = \frac{1}{K^n}.
\]
Since $\sigma_i$ are i.i.d. multinomial, then if $\pi$ is non-informative,
$\mathbf{\pi} = (1/K,\ldots,$ $1/K)$, the labels are not identifiable in the
posterior $\Pr(\sigma | A)$ either. In fact, non-identifiability issues occur
within a group of labels $\mathcal{I}$ whenever $\pi_i = \pi_j$ for all $i,j
\in \mathcal{I}$, but we discuss a non-informative $\pi$ for simplicity and
because that is a common modeling choice.

A common approach in latent class models to fix label non-identifiability is
to fix an arbitrary order in the parameters~\citep[Chapter~18]{gelman03},
e.g. $\gamma_{12} < \cdots < \gamma_{K-1,K}$. However, as \citet{nowicki01}
point out, this solution can lead to imperfect identification of the classes
if the parameters are close with high posterior probability; a major drawback
then is that parameters and labels can be interpreted incorrectly.
To address this problem, a label switching algorithm was proposed by
\citet{stephens00} in the context of MCMC sampling, but it is slow in
practice. Another approach is to simply focus on permutation-invariant
functions; in particular, when estimating $\sigma$, we can adopt a 
permutation-invariant loss, such as Binder's loss~\citep{binder78}. We discuss
such approach in more detail in Section~\ref{sec:postinf}. Next, we propose an
alternative, simpler procedure to remap labels and address
non-identifiability.

\subsection{Canonical Projection and Remapping Labels}
Let $L \doteq \{1, \ldots, K\}$ and 
$\mathcal{L} = \{\sigma \in L^n : N_k(\sigma) > 1, k = 1, \ldots, K\}$ be the
space of labels with positive prior probability.
If $\rho$ is any \emph{permutation} of the labels then
$\Pr(\sigma|A) = \Pr(\rho(\sigma)|A)$, where $(\rho(\sigma))_j =
\rho(\sigma_j)$ for $j=1,\ldots,n$. Non-identifiability here means that
$\Pr(\cdot \given A)$ is \emph{invariant} under $\rho$, and that $\sigma$ and
$\rho(\sigma)$ are $\Pr(\cdot|A)$-\emph{equivalent}, which we denote by
$\sigma \sim_P \rho(\sigma)$.
Moreover, we can partition $\mathcal{L}$ according to $\sim_P$: if $S$ is one
such partitioned subspace, then any $\sigma \in S$ is such that $\sigma$ is
not $\Pr(\cdot \given A)$-equivalent to any other label configuration in $S$.
To achieve label identifiability we anchor one such subspace as a
\emph{reference} space $Q$ and regard all other subspaces as permuted copies
of $Q$.

Let \textsf{ind}$(\sigma)$ be the vector with the first positions in
$\sigma$ where each label appears,
$
\mbox{\sf ind}(\sigma)_k \doteq
\min\{i : \sigma_i = k\},
$
and further define \textsf{ord}$(\sigma)$ as the vector with the order in
which the labels appear in $\sigma$,
\begin{equation}
\label{eq:ord}
\mbox{\sf ord}(\sigma)_k =
\sigma^{-1} \Big[ \mbox{\sf ind}(\sigma)_{(k)} \Big]
, \quad k \in L.
\end{equation}
Note that \textsf{ind}$(\sigma)_{(k)}$ is the $k$-th position in the ordered
vector \textsf{ind}$(\sigma)$. As an example, if
$\sigma = (2,2,3,1,3,4,2,1)$ with $K=4$ (and $n=8$) then
\textsf{ind}$(\sigma) = (4,1,3,6)$, ordered \textsf{ind}$(\sigma)$ is
$(1,3,4,6)$ and so \textsf{ord}$(\sigma) = (2,3,1,4)$.
To maintain identifiability we then simply constraint label assignments to
the subset of $\mathcal{L}$ where \textsf{ord}$(\cdot)$ is fixed. As a simple,
natural choice, let us restrict assignments to
$Q = \{\sigma : \mbox{\sf ord}(\sigma) = L\}$.
Note that any $\sigma$ can be mapped to its \emph{canonical} assignment
by
\begin{equation}
\label{eq:remap}
\rho(\sigma) \doteq \mbox{\sf ord}(\sigma)^{-1}(\sigma).
\end{equation}
Taking our previous example, $\sigma = (2, 2, 3, 1, 3, 4, 2, 1)$ would then be
mapped to $\rho(\sigma)=(1, 1, 2, 3, 2, 4, 1, 3)$.
The definitions of \textsf{ind} and \textsf{ord} can then be used to derive a
procedure that \emph{remaps} $\sigma$ to $\rho(\sigma)$; for completeness, we
list an algorithm that implements such remap procedure in
Appendix~\ref{sec:remap}.

Our proposed reference set above is also described by 
$Q = \{\sigma \in \mathcal{L}: \sigma = \rho(\sigma)\}$,
the \emph{quotient} space of $\mathcal{L}$ with respect to
\textsf{ord}, $\mathcal{L} / \text{\sf ord}$: any pair of label
configurations $\sigma_1$ and $\sigma_2$ such that $\rho(\sigma_1) =
\rho(\sigma_2)$ are identified to a single label $\rho(\sigma_1)$ in $Q$.
By constraining the labels to a reference quotient space we achieve not only
identifiability, but also make the labels interpretable: label $j$ marks
the $j$-th community to appear in the sequence of labels. As a consequence, we
are not restricted to estimating permutation-invariant functions of the
labels, as in the approach of \citet{nowicki01}, since now, for example,
$\Pr(\sigma_i = j \given A)$ is meaningful. As a particular application, we
derive a direct estimator of $\sigma$ based on Hamming loss in
Section~\ref{sec:postinf}; in the next section we discuss how the constraint
to $Q$ is implemented in practice.

%% file: postsamp.tex
\section{Posterior Sampling}
\label{sec:postsamp}

To sample from the joint posterior on $\sigma$, $\beta$ and $\pi$, we
use a Gibbs sampler \citep{geman84,robert99} that iteratively alternates
between sampling from
\[
[\sigma \given \gamma, \eta, \pi, A], \quad
[\pi \given \sigma, \gamma, \eta, A], \quad
[\gamma, \eta \given \sigma, \pi, A]
\]
until convergence. Next, we discuss how we obtain each conditional
distribution in closed form.

\subsection{Sampling $\sigma$ and $\pi$}
Let us start with the most relevant parameters: the labels $\sigma$. We can
sample a candidate, unconstrained assignment for actor $i$, $\sigma_i$,
conditional on all the other labels $\sigma_{[-i]}$, parameters $(\beta,
\pi)$, and data $A$ from a multinomial with probabilities:
\begin{multline}
\label{eq:condsigma}
\Pr(\sigma_i \given \sigma_{[-i]}, \beta, \pi, A) \propto
\pi_k \\
\qquad \prod_{j \ne i} \Big(
\text{logit}^{-1}(\gamma_{\sigma_i\sigma_j} + \eta_i + \eta_j)
\Big)^{A_{ij}}
\Big(
1 - \text{logit}^{-1}(\gamma_{\sigma_i\sigma_j} + \eta_i + \eta_j)
\Big)^{1 - A_{ij}} \\
=
\pi_k \prod_{j \ne i}
\frac{\exp\{A_{ij} (\gamma_{\sigma_i\sigma_j} + \eta_i + \eta_j)\}}%
{1 + \exp\{\gamma_{\sigma_i\sigma_j} + \eta_i + \eta_j\}}.
\end{multline}
To guarantee that parameters are identifiable, we reject the candidate
$\sigma$ if $N_k \le 1$ for any community $k$. Moreover, to keep the labels
identifiable, we remap $\sigma$ using the routine in Section~\ref{sec:label}
and remap $\gamma$ accordingly.

\begin{figure}[ht]
 \begin{center}
 \includegraphics[width=12cm]{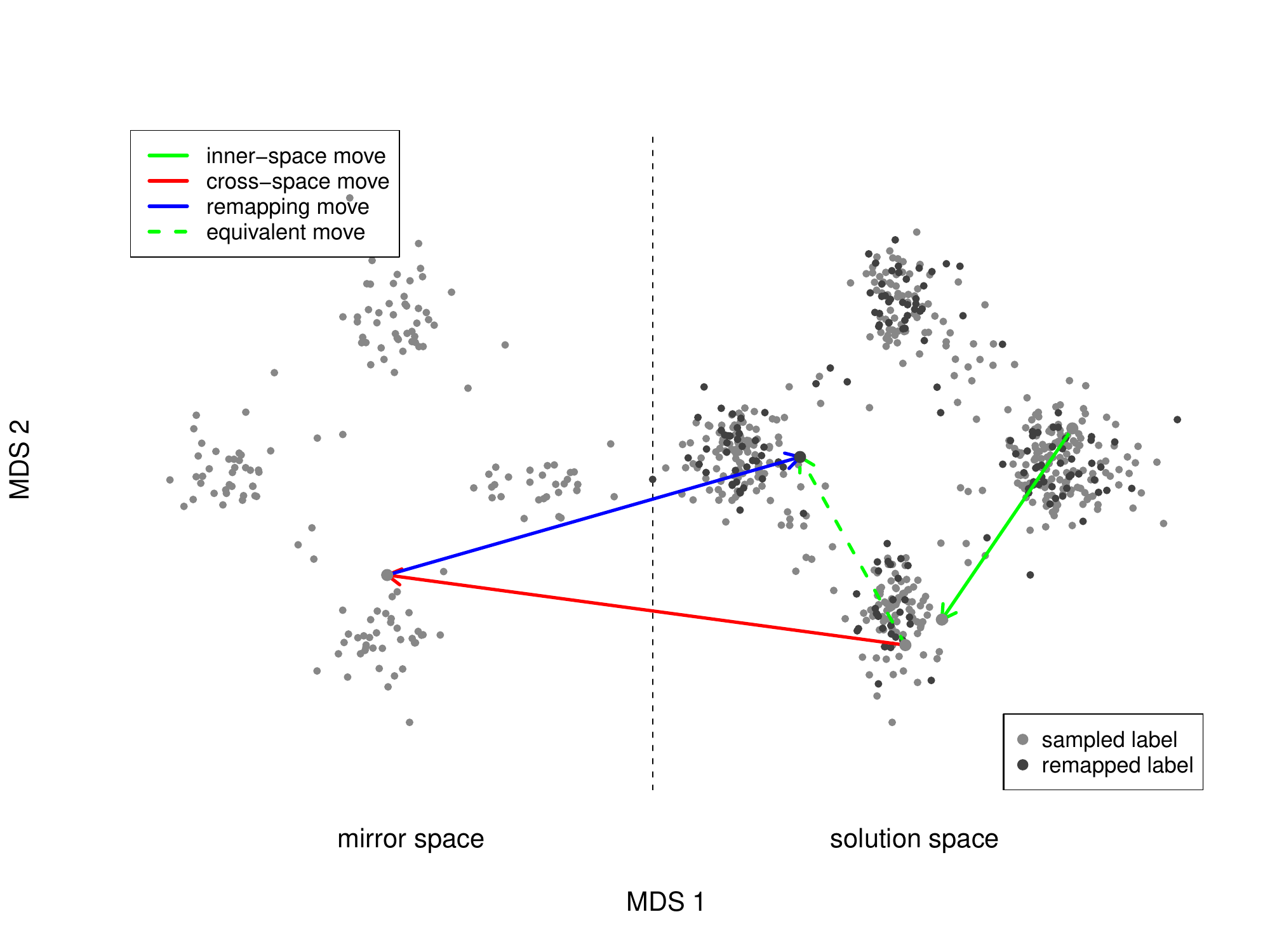}
 \end{center}
\caption{MDS representation of the two copies of the quotient space
$\mathcal{L} / \text{\sf ord}$ using posterior samples for the political blogs
example in Section~\ref{sec:exp}. Arrows are described in text.}
\label{fig:remap}
\end{figure}

As an example, consider the label samples obtained from running the Gibbs
sampler on the political blogs study in Section~\ref{sec:exp}. In
Figure~\ref{fig:remap} we plot a multidimensional scaling
[MDS~\citep{gower66}] representation of the samples. We have $K=2$
communities, and so $\mathcal{L}$ is partitioned into a reference quotient
space in the right and a ``mirrored'' space in the left; any point in the
mirrored space can be obtained by swapping labels $1$ and $2$ in the reference
space and vice-versa. The green arrow shows a valid sampling move
$\sigma^{(t)} \rightarrow\sigma^{(t+1)}$ at iteration $t$ that does not
require a remap, while the red arrow is an invalid move since it
crosses spaces. The blue arrow remaps $\sigma^{(t+1)}$ to
$\rho(\sigma^{(t+1)})$ in the reference space. The dashed green arrow
summarizes both operations.


For the nuisance parameter $\pi$ we summon conjugacy to obtain
\begin{equation}
\label{eq:condpi}
\pi \given \sigma, \theta, A \sim \mbox{\sf Dir}(\alpha + \mathbf{N}(\sigma)),
\end{equation}
where $\mathbf{N}(\sigma) = (N_1, \ldots, N_K)$ and $N_k$ are community
sizes.

\subsection{Sampling $\gamma$ and $\eta$}
Sampling $\beta$ conditional on $\sigma$, $\pi$, and data $A$ is more
challenging since the logistic likelihood in~\eqref{eq:logitsbm} does not
specify a closed form distribution. However, if we explore a data augmentation
strategy by introducing latent variables
$\omega = (\omega_{ij})_{i < j: i,j \in \{1, \dots, n\}}$ from a P\'olya-Gamma
distribution, then the above conditional distribution of
$\beta$ given $\omega$ is now available in closed form~\citep{polson12}. More
specifically, if
$\omega_{ij} \given \sigma, \beta \sim
\text{\sf PG}(1, {x_{ij}(\sigma)^\top} \beta)$, then
\[
\beta \given \omega, \sigma, A \sim I[\gamma \leq 0] \cdot N(m, V)
\]
where, with $\Omega=\text{Diag}(\omega_{ij})$ and latent weighted
responses $z_{ij}=(A_{ij}-1/2)\omega_{ij}^{-1}$,
\begin{equation}
\label{eq:mv}
V = \Big(X^\top\Omega X+\frac{1}{\tau^2} I_{n+\binom{K}{2}}\Big)^{-1}
\quad \text{and} \quad
m = VX^\top\Omega z.
\end{equation}

The assortativity constraint $\gamma \le 0$ in the $\beta$ prior is clearly
also present in the conditional posterior, and so we can use a simple
rejection sampling step for the truncated normal: sample from unconstrained
marginals $N(m, V)$ and accept only if $\gamma \leq 0$. However,
since
\[
\beta = \left[ \begin{array}{c} \gamma \\ \eta \end{array} \right]
\,\Bigg|\, \omega, \sigma, A
\sim N\Bigg( m = \left[ \begin{array}{c}
m_{\gamma} \\
m_{\eta} \end{array} \right], V = \left[ \begin{array}{cc}
V_{\gamma} & V_{\gamma \eta} \\
V_{\eta \gamma } & V_{\eta}  \end{array} \right] \Bigg),
\]
we can adopt a more efficient way of sampling $\beta$ by first sampling
$\eta$ marginally,
\begin{equation}
\label{eq:condeta}
\eta \given \omega, \sigma, A \sim N(m_{\eta}, V_{\eta}),
\end{equation}
and then sampling
\begin{equation}
\label{eq:condgamma}
\gamma \given \eta, \omega, \sigma, A \sim I(\gamma \le 0) \cdot N(
m_{\gamma} + V_{\gamma \eta}V_{\eta}^{-1}(\eta - m_{\eta}),
V_{\gamma} - V_{\gamma \eta}V_{\eta}^{-1}V_{\eta\gamma})
\end{equation}
from a truncated normal. In practice, we compute the Schur complement of
$V_{\eta}$, $V_{\gamma} - V_{\gamma \eta}V_{\eta}^{-1}V_{\eta\gamma}$, using
the SWEEP operator~\citep{goodnight79}.

\subsection{Gibbs sampler}
To summarize, after setting initial parameters $\sigma$, $\beta$ and $\pi$
arbitrarily, we then iterate until convergence the following Gibbs sampling
steps:

\begin{enumerate}
  \item Sample $\sigma \given \beta, \pi, A$: for each node $i$,
  \begin{enumerate}
    \item Sample $\sigma_i \given \sigma_{[-i]}, \beta, A$ from a
    multinomial distribution as in~\eqref{eq:condsigma}. If $N_k(\sigma) < 2$
    for some community $k$, reject and keep the previous value of $\sigma_i$.
    \item Remap $\sigma$ using the procedure in Section~\ref{sec:label}.
  \end{enumerate}

  \item Sample $\pi \given \sigma, \beta, A$ from the Dirichlet distribution
  in~\eqref{eq:condpi}.

  \item Sample $\beta \given \sigma, \pi, A$:
  \begin{enumerate}
    \item Sample $\omega \given \sigma, \beta, \pi, A$: for each pair $i < j$,
    $\omega_{ij} \given \sigma, \beta \sim \text{\sf PG}(1,
    x_{ij}(\sigma)^\top \beta)$.

    \item Sample $\beta \given \sigma, \pi,\omega, A$: compute $m$ and $V$ as
    in~\eqref{eq:mv}, sample $\eta$ marginally as in~\eqref{eq:condeta}, and
    then sample $\gamma \given \eta$ from a truncated multivariate normal
    distribution as in~\eqref{eq:condgamma}.
  \end{enumerate}

\end{enumerate}

To speed up convergence and improve precision, we set the initial $\sigma$ to
be an approximate posterior mode obtained from a greedy optimization version
of the above routine, similar to a gradient cyclic descent method. The main
changes are:

\begin{enumerate}
\item In Step~1.a we take $\sigma_i$ to be the mode of $\sigma_i \given
\sigma_{[-i]}, \beta, A$ (but we might still reject $\sigma_i$ if
$N_k(\sigma) < 2$ for some $k$ and remap $\sigma$ in Step~1.b.)

\item In Step~2, we take $\pi$ to be the mode of the Dirichlet distribution
in~\eqref{eq:condpi}.

\item Step~3 is substituted by a regularized iterative reweighted
least-squares (IRLS) step. IRLS is usual when fitting logistic regression
models~\citep{mccullagh89}. At the $t$-th iteration we define
$\mu_{ij} = \text{logit}^{-1}(x_{ij}(\sigma)^\top \beta^{(t)})$ and 
$W = \text{Diag}(\mu_{ij} (1 - \mu_{ij}))$ to obtain the update
\[
V = \Big(X^\top W X + \frac{1}{\tau^2} I_{n+\binom{K}{2}}\Big)^{-1}
\quad \text{and} \quad
\beta^{(t+1)} = VX^\top W z^{(t)}
\]
where $z^{(t)} = X \beta^{(t)} + W^{-1}(y - \mu)$ is now the ``working
response''. To guarantee that the community constraints $\gamma \le 0$ are
met, we use an active-set method~\citep[Chapter~16]{nocedal06}.

\end{enumerate}

Since we expect the posterior space to be multimodal, we adopt a strategy
similar to~\citet{karrernewman11} and sample multiple starting points for
$\sigma$ according to its prior distribution and then obtain approximate
posterior modes for each simulation. We elect the best approximate mode over
all simulations as the starting point for the Gibbs sampler, which is then run
until convergence to more thoroughly explore the posterior space.
For convenience, the Gibbs sampler and its optimization version are
implemented in the \texttt{R} package \texttt{sbmlogit}, available as
supplementary material.


%% file: postinf.tex
\section{Posterior Inference}
\label{sec:postinf}
The usual estimator for label assignment is the maximum \emph{a posteriori}
(MAP) estimator,
\[
\hat{\sigma}_M = \argmin_{\tilde{\sigma} \in \{1, \ldots, K\}^n}
\Exp_{\sigma \given A} \big[ I(\tilde{\sigma} \ne \sigma) \big]
= \argmax_{\tilde{\sigma} \in \{1, \ldots, K\}^n}
\Pr(\sigma = \tilde{\sigma} \given A),
\]
which, albeit based on a zero-one loss function \citep{besag86}, has the
advantage of being invariant to label permutations. However, given the
flexibility in our model due to the hierarchical levels, the posterior space
is often complex and so the MAP might fail to capture the variability and
might focus on sharp peaks that gather a small amount of posterior mass around
them.

Another estimator for label assignment arises from minimizing Binder's loss
$B$ \citep{binder78, binder81},
\begin{equation}
\label{eq:binder}
\hat{\sigma}_B = \argmin_{\tilde{\sigma} \in \{1, \ldots, K\}^n}
\Exp_{\sigma \given A} \big[ B(\tilde{\sigma}, {\sigma}) \big],
\end{equation}
where
\[
B(\tilde{\sigma}, {\sigma}) = \sum_{i<j} I(\tilde{\sigma}_i \ne
\tilde{\sigma_j})I(\sigma_i = \sigma_j) + I(\tilde{\sigma}_i =
\tilde{\sigma_j})I(\sigma_i \ne \sigma_j).
\]

The advantage of Binder's loss is that since it penalizes pairs of nodes it is
invariant to label permutations---that is,
$B(\tilde{\sigma}, \sigma) = B(\tilde{\sigma}, \phi(\sigma))
= B(\phi(\tilde{\sigma}), \sigma)$ for any permutation $\phi$.
However, \citet{laugreen07} have shown that minimizing Binder's loss is
equivalent to binary integer programming, which is an NP-hard problem.
Moreover, as \citet{fritsch09} point out, even the approximated solution given
by \citet{laugreen07} is only feasible when the dataset is of moderate size.

In contrast, when compared to MAP inference, centroid
estimation~\citep{carvalho08} offers a better representative of the space
since it arises from a loss function that is more refined:
\[
\hat{\sigma}_H = \argmin_{\tilde{\sigma} \in \{1, \ldots, K\}^n}
\Exp_{\sigma \given A} \big[ H(\tilde{\sigma}, \sigma) \big],
\]
where $H$ is \emph{Hamming} distance, $H(\tilde{\sigma}, \sigma) =
\sum_{i=1}^n I(\tilde{\sigma}_i \ne \sigma_i)$. The centroid estimator also
identifies the median probability model, and thus is known to offer better
predictive resolution that the MAP estimator~\citep{barbieri04}. 
However, Hamming loss is only invariant to double label permutations but not
to single label permutations, i.e.,
$H(\tilde{\sigma}, \sigma) = H(\phi(\tilde{\sigma}), \phi(\sigma))$ but
it is not necessarily true that 
$H(\tilde{\sigma}, \sigma) = H(\phi(\tilde{\sigma}), \sigma)$ or
$H(\tilde{\sigma}, \sigma) = H(\tilde{\sigma}, \phi(\sigma))$, and
thus, in order for Hamming loss to be meaningful for estimation when the
labels are non-identifiable we need to account for label aliasing. We then
redefine the centroid estimator to depend on a specific permutation, for
instance the canonical permutation $\rho$ in~\eqref{eq:remap},
\[
\hat{\sigma}_C = \rho \Bigg(
\argmin_{\tilde{\sigma} \in \{1, \ldots, K\}^n}
\Exp_{\sigma \given A} \big[ H(\tilde{\sigma}, \rho(\sigma)) \big]
\Bigg).
\]
This remapped centroid estimator considers only one version of the posterior
space, namely the reference quotient space $\mathcal{L} / \text{\sf ord}$ with
\textsf{ord} in~\eqref{eq:ord}. The main advantage of this new estimator is to
allow the following characterization (see Appendix~\ref{sec:proofcent} for the
proof):

\begin{theorem}
\label{thm:centroid}
The centroid estimator $\hat{\sigma}_C$ is a mapped \emph{consensus}
estimator: if $\Pr^*(\sigma \given A)$ is the induced posterior probability of
$\sigma \in \mathcal{L} \,/\, \mbox{\sf ord}$ and
\[
(\hat{\sigma}^*)_i = \argmax_{k \in \{1, \ldots, K\}}
\Pr^*(\sigma_i = k \given A)
\]
then $\hat{\sigma}_C = \rho(\hat{\sigma}^*)$.
\end{theorem}

In practice, we estimate
\[
\hat{\Pr}^*(\sigma_i = k \given A) \approx
\frac{1}{N}\sum_{t=1}^N I(\sigma^{(t)}_i = k)
\]
using the realizations from the Gibbs sampler presented in
Section~\ref{sec:postsamp} to define $\hat{\sigma}_C$. Since we only need to
elect, for each actor, the most likely label, obtaining the centroid estimator
is much simpler computationally than MAP and Binder estimation. Note that due
to the remap step when sampling $\sigma \given \theta, A$, we are always
constrained to the quotient space $\mathcal{L} / \mbox{\sf ord}$ and
identifying label realizations under $\rho$, and thus really approximating
$\Pr^*(\sigma \given A)$.

\subsection{Relating Binder and Centroid Estimators}
We start by noting that if we define an extended \emph{matched} map
$M(\sigma) = \{I(\sigma_i = \sigma_j)\}_{1 \leq i < j \leq n}$
that makes pairwise comparisons among labels in $\sigma$, then Binder and
Hamming losses are related through
$B(\tilde{\sigma}, \sigma) = H(M(\tilde{\sigma}), M(\sigma))$ and so Binder's
estimator in~\eqref{eq:binder} is also a centroid estimator in the extended
matched space $M(\mathcal{L})$.

Back to the original space $\mathcal{L}$ of labels, we observe that, in
practice, the Binder and centroid estimators are often close (in either loss.)
To explain these observations, we need the next result relating Binder and
Hamming losses (the proof can be found in Appendix~\ref{sec:binder}):

\begin{theorem}
\label{thm:binder}
For any pair of label assignments $\tilde{\sigma}$ and ${\sigma}$, Binder
loss is bounded by Hamming loss through
\begin{equation}
\label{eq:binderbound}
B(\tilde{\sigma}, {\sigma}) \leq H(\tilde{\sigma}, {\sigma})
\Big(n - \frac{1}{2}H(\tilde{\sigma}, {\sigma})\Big).
\end{equation}
Moreover, if $K=2$ then
$B(\tilde{\sigma}, \sigma) =
H(\tilde{\sigma}, \sigma) (n - H(\tilde{\sigma}, \sigma))$.
\end{theorem}

From~\eqref{eq:binderbound} we see that Binder's loss can be approximately
linearly bounded by Hamming loss when the Hamming distance between
$\tilde{\sigma}$ and ${\sigma}$ is small. Thus, when the marginal posterior
distribution on $\sigma$ has a compact cluster of label configurations with
high posterior mass we expect this cluster to contain the centroid estimator
and also, according to~\eqref{eq:binderbound}, the Binder estimator since
minimizing the posterior expected Hamming loss is approximately equivalent to
minimizing the posterior expected Binder loss in this case.
In the next section we run experiments on simulated datasets and observe that 
the two estimators are often close and show similar performance for simple
networks (check, for instance, Figure~\ref{fig:bm}.)

%% file: experiment-carvalho.tex
\section{Experimental Results}
\label{sec:exp}
In this section, we demonstrate the performance of the centroid estimator and
compare it to Binder estimator under our model and to KN
estimator~\citep{karrernewman11}, Fast-Greedy (FG)
estimator~\citep{Clauset04}, Multi-Level (ML) estimator~\citep{Blondel08},
Walktrap (WT) estimator~\citep{Pons04} and Label Propagation (LP)
estimator~\citep{Raghavan07} through an empirical study and two case
studies. In the case studies we run repeated experiments on the same dataset
and obtain the error rates of the estimators mentioned above when
compared to known or \emph{bona fide} ground truth references. To compare
those estimators, we define a $q$-\emph{error interval} as the interval with
endpoints being the $q/2$ and $1-q/2$ quantiles of the error rates.

Before discussing the experimental results, we present two illustrative
examples next.

\subsection{Illustrative Examples}
\label{ssec:examples}
Even though the models reviewed above are flexible enough to identify social
block structure, they might fail to actually recognize communities. We now
show two simple examples to demonstrate how this happens, and compare our
proposed solution to the results from applying Karrer and Newman's (KN)
popular degree-corrected SBM~\citep{karrernewman11}. 

\begin{figure}[ht]
\begin{center}
\includegraphics[width=.8\textwidth]{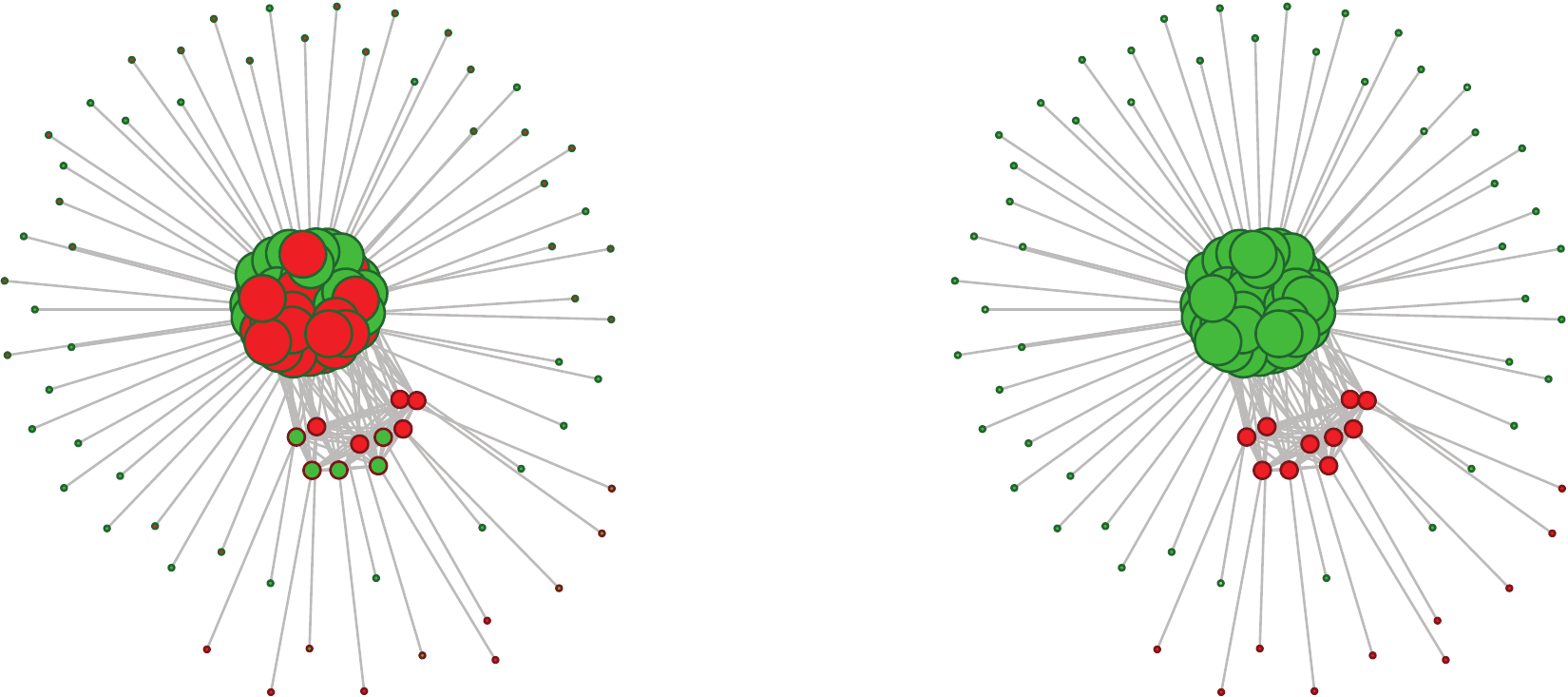}
\end{center}
\caption{Spike network, $n_1=10$, $r=5$. Node sizes are proportional to
degree; node colors (red/green) represent groups in KN estimator (left) and
our estimator (right.) Node borders mark the reference.}
\label{fig:spikesol}
\end{figure}

The first dataset is a synthetic network, denoted as the ``spike''
dataset, which we intentionally designed to show that degree correction is
not sufficient to elicit communities. The network considered is split into
$K=2$ communities. The first community contains $2n_1$ nodes with $n_1$ of
them being strongly connected as a complete graph $K_{n_1}$ (a ``kernel'') and
having a one-to-one connection with the remaining $n_1$ nodes (a ``crown'').
The other community is formed in a similar way, but with a complete $K_{rn_1}$
kernel connected to a $rn_1$ crown, totalling $2rn_1$ nodes (see
Figure~\ref{fig:spikesol}). We add some between-community edges in such a way
that each node from the complete graph $K_{n_1}$ in the first community is
connected to $r$ nodes from the complete graph $K_{rn_1}$ in the second
community.

We note that this network can still be characterized as having a
community behavior since the edge density between communities is smaller than
the density within communities. Moreover, due to the crowns, we also need to
account for degree heterogeneity in each community.
Let us then consider the case when $n_1=10$ and $r=5$.
Figure~\ref{fig:spikesol} compares the KN estimator and our estimator. The
kernel-crown structure of both communities is not reflected in KN estimator;
moreover, there are more edges between groups than within groups, which is not
prescribed by community behavior.


\begin{figure}[ht]
\begin{center}
\includegraphics[width=.8\textwidth]{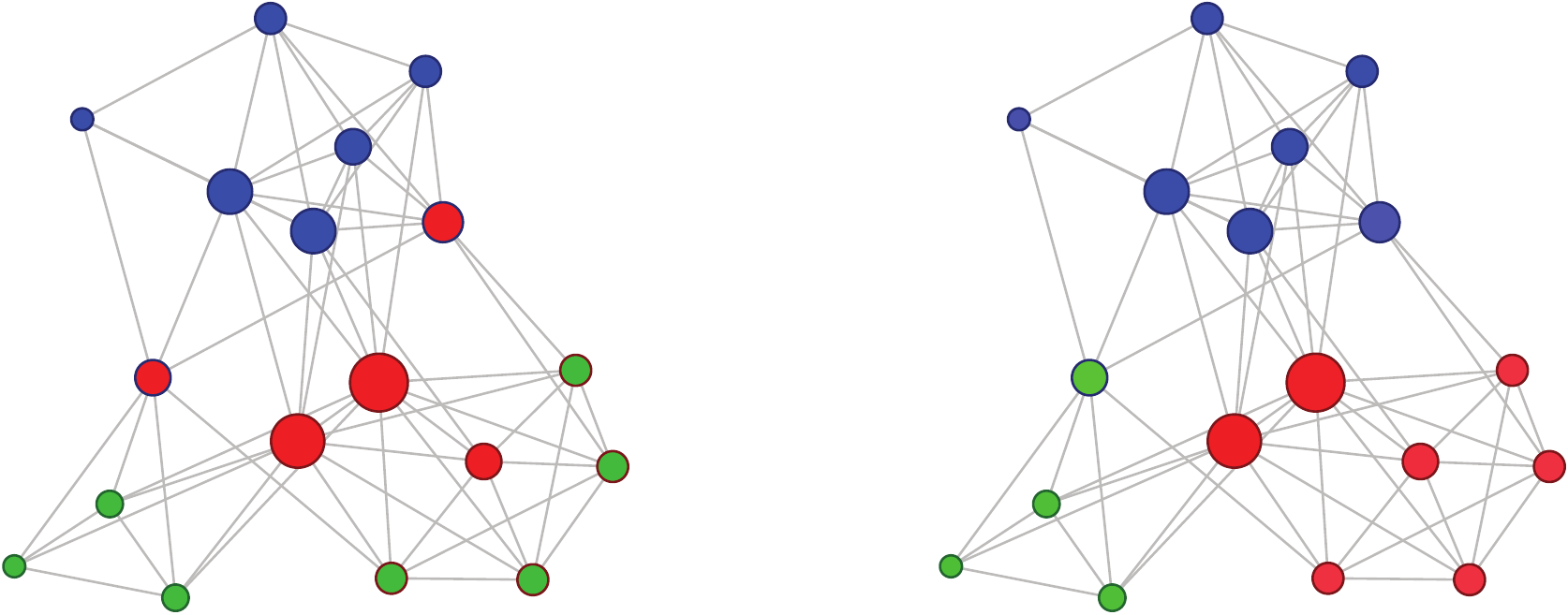}
\end{center}
\caption{Sampson network at $T_4$, $n=18$. Node sizes are proportional to
degree; node colors mark KN estimator (left) and our estimator
(right). Node borders mark the reference.}
\label{fig:solst4}
\end{figure}

We observe that degree correction is not enough to correctly capture the
community structure in the synthetic network that we designed.
However, similar results are also observed in some real-world datasets.
Consider, for example, the ``sampson'' network reported by~\citet{sampson68}
at time point $T_4$ among a group of $18$ trainee monks at a New England
monastery. Four types of relations---affection, esteem, influence, and
sanctioning---between the monks are collected. In this network, each node
represents a monk in the monastery, and two nodes are considered to be
connected if they considered each other as being in at least one of the four
relations when asked by Sampson. Sampson reported a partition of trainee monks
into three communities ($K=3$): Young Turks, Loyal Opposition and Outcasts.
Figure~\ref{fig:solst4} compares KN estimator to our estimator and shows a
similar pattern where within group connections are sparser than between group
connections according to the KN estimate; in particular, there are more edges
between the red and green communities than within the green community.
In fact, the KN partition does not agree with any well-accepted reference.


\subsection{Empirical Study}
First, we evaluate our estimator on simulated data\-sets with known references.
The networks are generated from a class of benchmark graphs that account for
heterogeneities in node degree distributions and community sizes
\citet{andrea08}. The model used in the simulation considers the following
parameters: both degree distribution and the community sizes are assumed to
follow power law distributions with exponents $a$ and $b$, respectively; each
network consists of $n$ nodes and has average degree $\langle k \rangle$; and
mixing parameter $\mu$ represents the proportion of the between-community
edges.

\begin{figure}[ht]
\begin{center}
\begin{tabular}{cc}
\includegraphics[width=.4\textwidth]{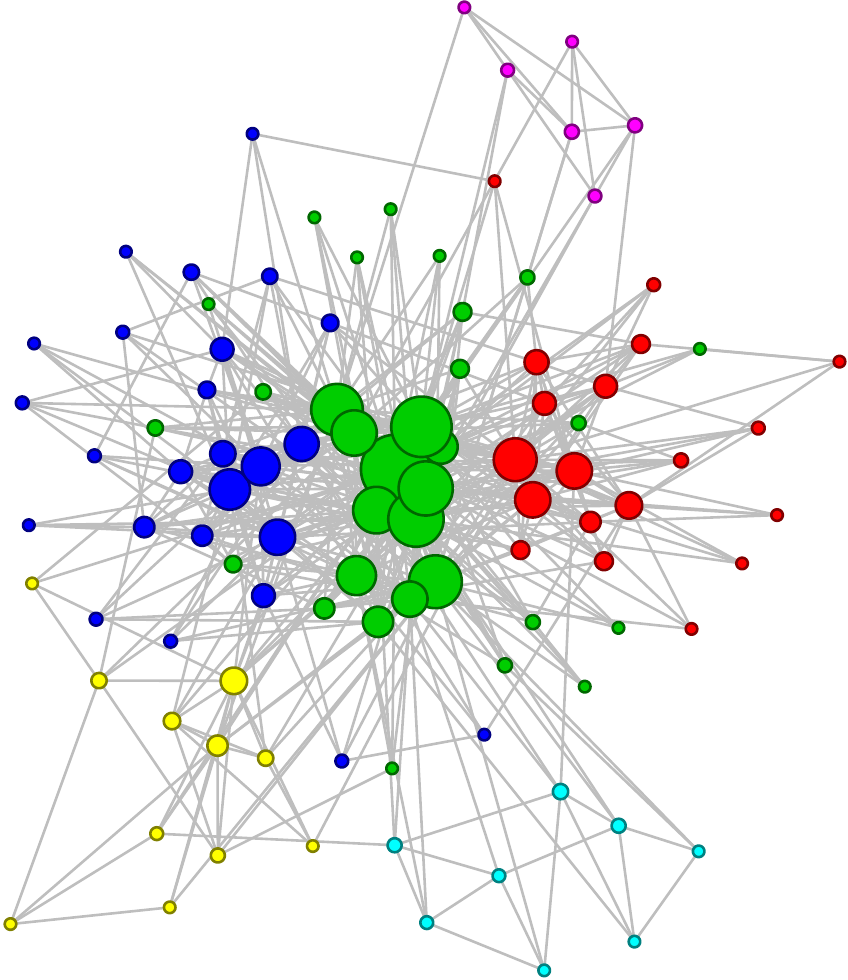} &
\includegraphics[width=.5\textwidth]{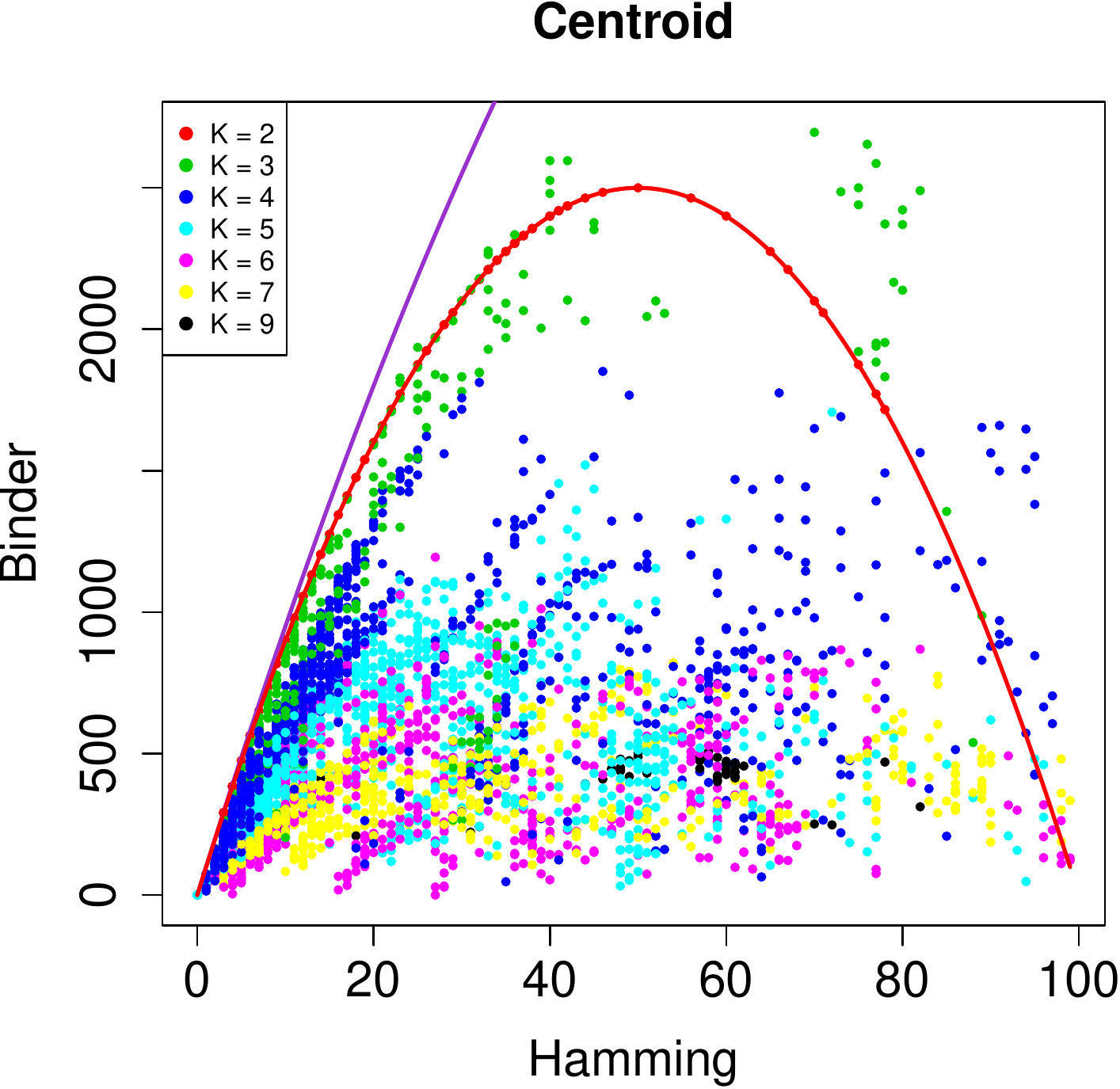}
\end{tabular}
\end{center}
\caption{Left: one realization of the benchmark networks with $n = 100$ nodes,
$a=2$, $b=1$, $\mu = 0.4$, and $\langle k \rangle=10$. Right: Binder
loss against Hamming loss tested on 50 graph realizations of such benchmark
networks. Colors mark different values of $K$. Lines correspond to the upper
bound in~\eqref{eq:binderbound} for $K > 2$ and $K = 2$.}
\label{fig:eghb}
\end{figure}

\begin{figure}[ht]
\begin{center}
\includegraphics[width=.9\textwidth]{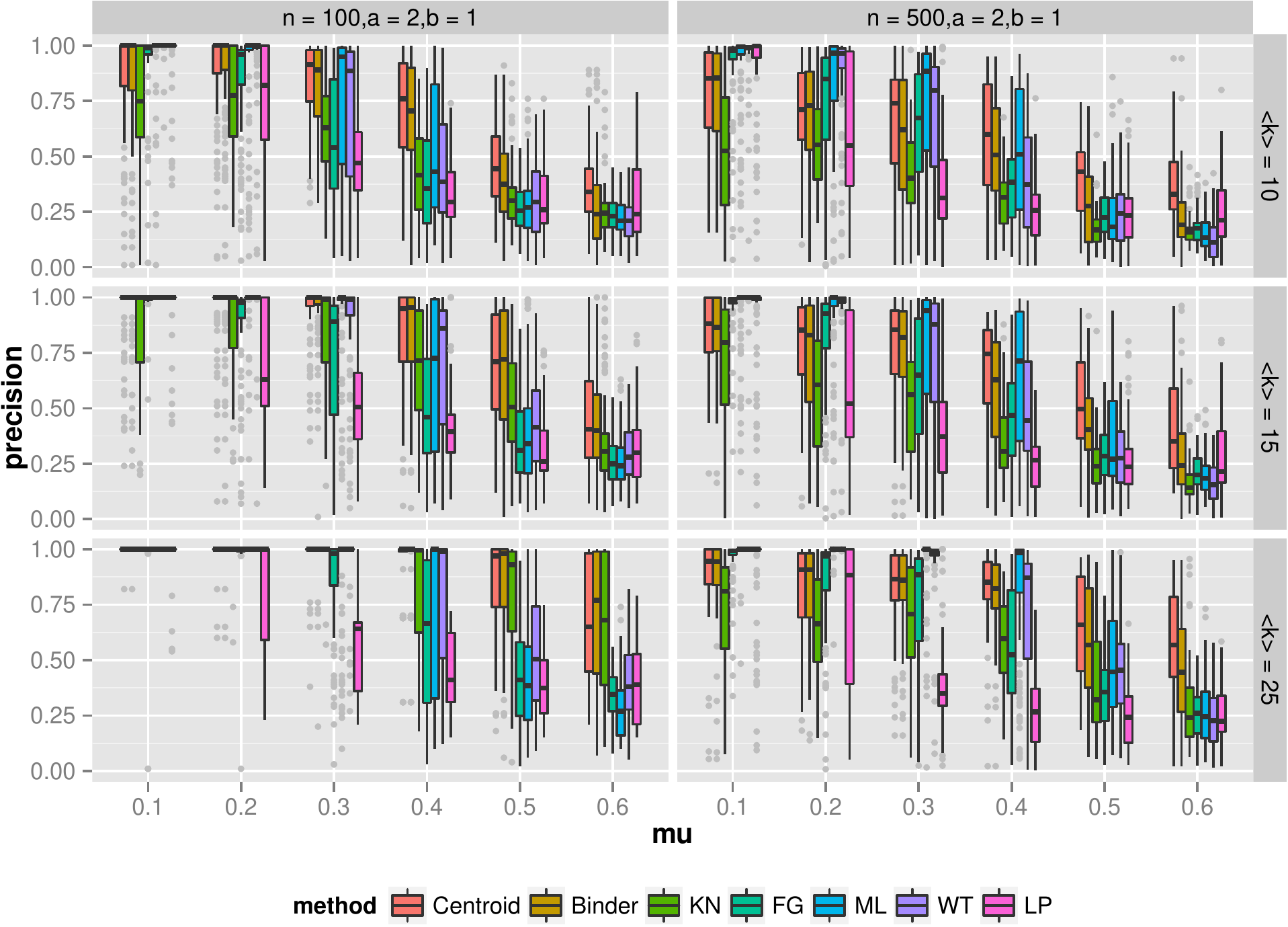}
\end{center}
\caption{Benchmark networks of $n = 100$ and $500$ nodes, with different
combinations of the exponents $a$, $b$ and the average degree $\langle k
\rangle$ are used. Each boxplot corresponds to the precision of the estimator
over 100 and 50 graph realizations for $n = 100$ and $n = 500$, respectively.
}
\label{fig:bm}
\end{figure}

We simulate $100$ networks for each combination of $n=(100,500)$, $a=(2,3)$,
$b=(1,2)$, and $\mu=(0.1,0.2,0.3,0.4,0.5,0.6)$.
Figure~\ref{fig:eghb} shows one realization of the benchmark networks as an
example. The precision of the centroid, Binder, KN, FG, ML, WT and LP estimators
are summarized in Figure~\ref{fig:bm}. We observe from the figure that
the centroid estimator yields smaller error rates than the KN estimator
while performing slightly better than Binder estimator in terms of mean error
rate. Besides, the centroid estimator performs comparably to FG, ML, WT and LP
estimators when the mixing parameter $\mu$ is small but outperforms these four
estimators to a large extent when the mixing parameter or the average degree
is relatively large. Not surprisingly, all estimators perform worse as the
mixing parameter $\mu$ increases (so that the communities are defined in a
weaker sense) or the average degree $\langle k \rangle$ decreases.
Similar results are found under other different combinations of $(a, b,
\langle k \rangle)$, as shown in Figure~\ref{fig:bm100} and
Figure~\ref{fig:bm500} in the Appendix.

\subsection{Case Study}
Next, we evaluate our estimator for community detection on two real-world
network datasets.

\subsubsection{Political blogs}
The first case study is the political blogs network \citep{adamic05},
which is a medium real-world network containing over one thousand nodes. In
this network, each node is a blog over the period of two months preceding the
U.S. Presidential Election of 2004, and two nodes are considered to be
connected if they referred to one another and there was overlap in the topics
they discussed. The network is known to be split into two communities ($K=2$),
liberals and conservatives, and has $n=1,\!222$ nodes after isolated nodes are
removed. It is expected that blogs in favor of the same party are more likely
to be linked and discussing the same topics than those in favor of different
parties, which corroborates a community behavior.

\begin{figure}[ht]
\begin{center}
\includegraphics[width=.9\textwidth]{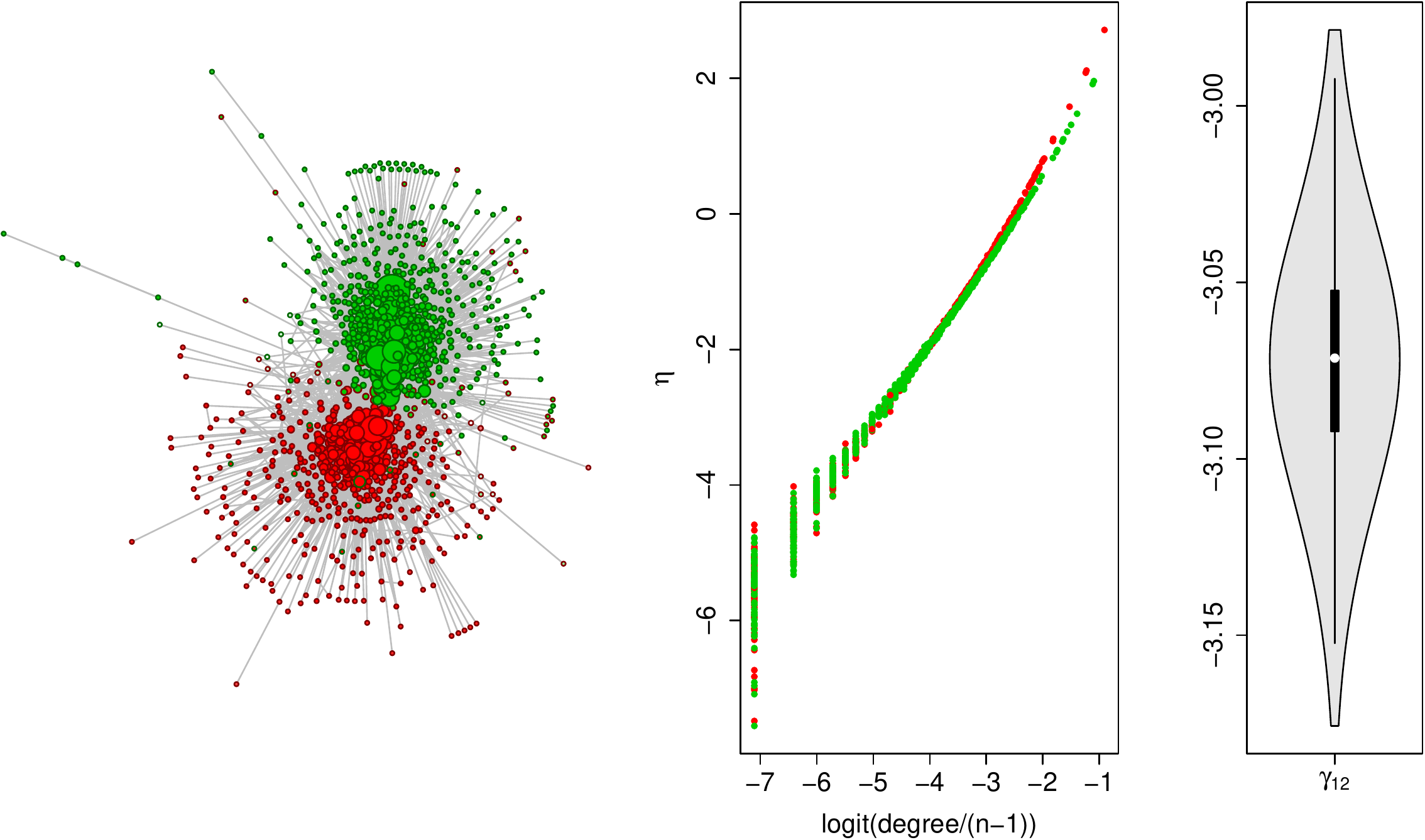}
\end{center}
\caption{Political blogs network. Left: Node sizes are proportional to
degree; node colors signal the centroid estimators (red/green). Node color
intensities are proportional to $\hat{\Pr}^*(\sigma_i \given A)$ and node
borders mark the reference. Middle: $\eta_i$ on
logit$(\mbox{degree}_i/(n-1))$ for each node $i$; color for each node $i$
represents ${(\hat{\sigma}_C)}_i$. Right: estimated posterior distribution for
$\gamma_{12}$.}
\label{fig:sol1}
\end{figure}

The centroid estimator, depicted in the leftmost panel in
Figure~\ref{fig:sol1}, agrees well with the reference of this network.
We estimate each $\eta_i$ for node $i$ by its estimated posterior mean using
the converged samples and plot the estimated $\eta_i$ against the logit
normalized degree of node $i$ in the middle panel of Figure~\ref{fig:sol1}.
There is a positive linear relationship between $\eta_i$ and the logit of the
normalized degrees, indicating that the expected degree, thus the probability
of having an edge, is positively related to the observed degree of the node.
If there is a community effect, that is, if the network can be better
explained by partitioning nodes into two different communities, then
$\gamma_{12}$ is expected to be significantly negative. The rightmost panel in
Figure~\ref{fig:sol1} shows the estimated posterior distribution of
$\gamma_{12}$.  An estimated 95\% credible interval for $\gamma$ is
$[-3.16, -2.99]$, which shows a clear deviation from $0$ and
thus indicates a strong community effect in the network.

We further compare the centroid estimator with two other
estimators, Binder and KN, as in the previous section. The estimated $90\%$
error intervals for the centroid, Binder, and KN estimators are
$[0.053, 0.054]$, $[0.053, 0.054]$, and $[0.045, 0.051]$, respectively.
In general, the three estimators perform equally well while the KN estimator
yields a slightly smaller error rate on average.

\subsubsection{Political books}
Finally, we pick the political books dataset compiled by Valdis Krebs
(unpublished). This is a network of political books sold by the on-line
bookseller Amazon around the time of the US presidential election in 2004. The
network is split into three communities: liberal, neutral, or conservative. An
edge between two books represents frequent co-purchasing by the same buyers.
We also use weakly-informative priors and run multiple chains. The estimated
$90\%$ error intervals for the centroid, Binder, and KN estimators are
$[0.167, 0.175]$, $[0.167, 0.175]$, and $[0.171, 0.171]$, respectively.
The reason why we observe large error rates under all estimation procedure
analyzed here might be that the reference provided by Valdis Krebs is not that
reliable, or that misclassified books appeal to buyers who purchase books from
all three political opinions. Most of the misclassified nodes are in the
neutral (red) community.

\begin{figure}[ht]
\begin{center}
\includegraphics[width=.9\textwidth]{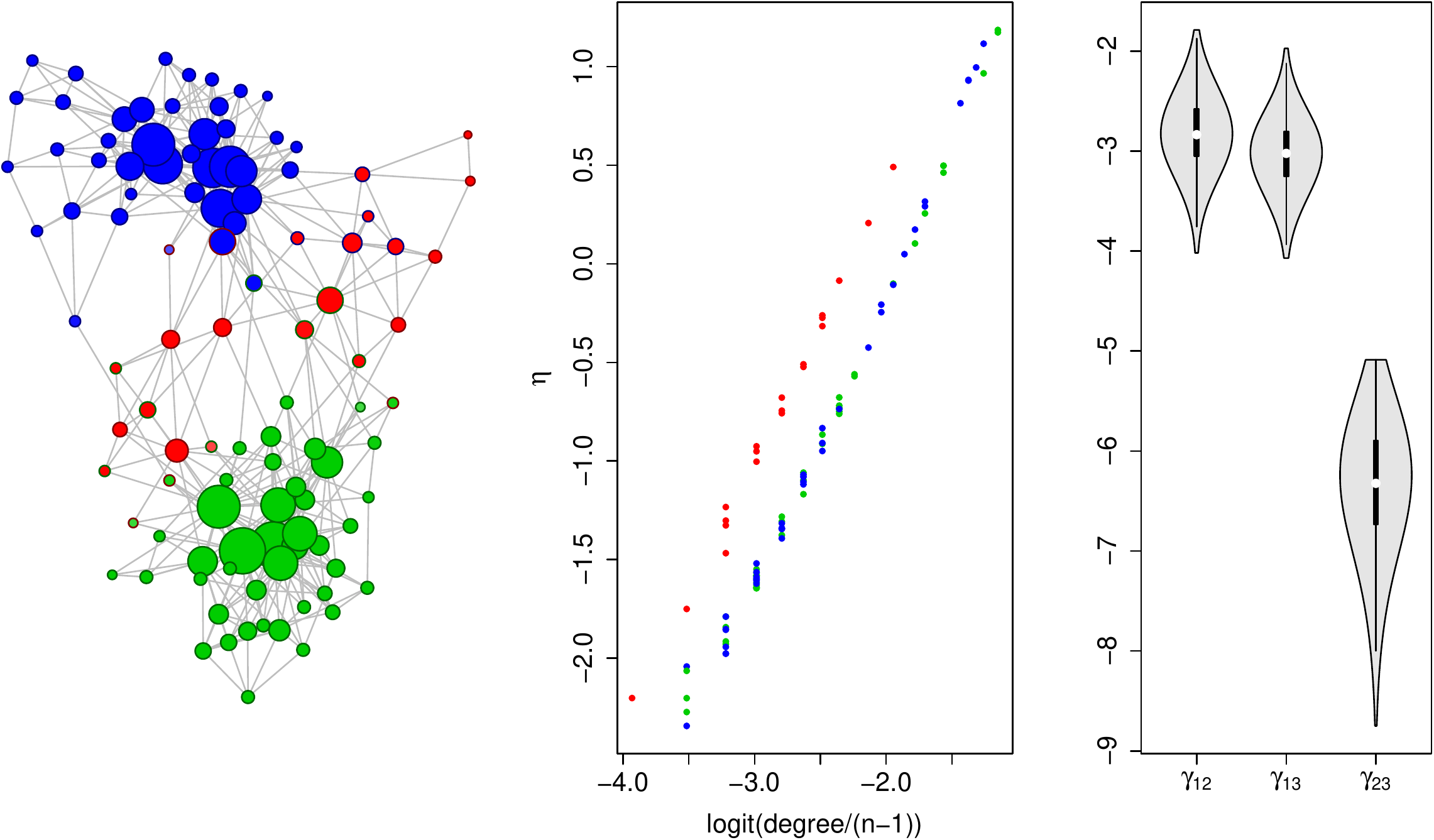}
\end{center}
\caption{Political books network. Left: node sizes are proportional to degree;
node colors signal the centroid estimators. Node color intensities are
proportional to $\hat{\Pr}^*(\sigma_i \given A)$ and node borders mark the
reference. Middle: $\eta_i$ on logit$(\mbox{degree}_i/(n-1))$ for each node
$i$; color for each node $i$ represents ${(\hat{\sigma}_C)}_i$.
Right: estimated posterior distribution for $\gamma$.}
\label{fig:polbookssol}
\end{figure}

Figure~\ref{fig:polbookssol} shows the centroid estimator of the political
books network in the right panel. The communities corresponding to liberal
(blue) and conservative (green) are clearly separated by the neutral (red)
community and agree with the reference well.
The middle panel plots estimated $\eta_i$ against normalized degrees in logit
scale; it is evident that the in-between red community has a different
intercept for $\eta$, indicating that it is less connected.
The right panel shows estimated marginal posterior distributions for $\gamma$.
Not surprisingly, $\gamma_{23} < \gamma_{12}$ and $\gamma_{23} < \gamma_{13}$
with high posterior probability since communities $2$ (green) and $3$ (blue)
are separated by community $1$ (red) and so do not share many edges.

%% file: discussion-carvalho.tex
\section{Discussion}
\label{sec:disc}
In this paper we have proposed a Bayesian model based on degree-corrected
stochastic blockmodels that is tailored for community detection. More
specifically, our model is flexible due to its hierarchical structure and aims
to capture the gregarious community behavior by requiring, through prior
specification, that the probability of within-community associations to be no
smaller than the probability of between-community associations.
Moreover, we argue that the model is a better representative of
assortatively mixing networks with binary data coding the associations instead
of frequency counts, since we model binary observations using a suitable
logistic regression with parameters for within and between-community
probabilities of association. We devise a Gibbs sampler to obtain posterior
samples and exploit a latent variable formulation to yield closed-form
conditionals.

We formally address label identifiability by restricting label configurations
to a canonical reference subspace, and propose a remap procedure to implement
this constraint in practice. As a consequence, labels are interpretable and we
are able to estimate any function of the labels as opposed to previous
approaches that were restricted to permutation-invariant functions.
In particular, we propose a novel remapped centroid estimator to infer
community assignments. We contend that while the model can arguably represent
the data well, the posterior space can be complex and a bad estimator can
spoil the analysis; it is then imperative to adopt an estimator that arises
from a principled and refined loss function and thus better summarizes the
posterior space. Our proposed remapped centroid estimator is more similar to
a posterior mean, and thus, while considering the whole posterior distribution
in the space of remapped label assignments, tends to situate itself in regions
of high concentration of posterior mass. From a practical point of view, we
show that the proposed estimator performs better than MAP and Binder
estimators and achieves lower misclassification rates.

If the posterior space is multimodal then a single point estimator has
difficulty in representing the space, and the centroid estimator is not immune
to this problem. We intend to further extend the proposed estimation procedure
to account for multiple modes by exploring \emph{conditional} estimators on
partitions of the space. While this can be done empirically by clustering
posterior samples, we will pursue a more principled way of identifying
partitions.
As simple extensions to the proposed model, we also intend to incorporate
parameters for node attributes and to generalize the formulation to account
for count, categorical, and ordinal data. Other directions for future work,
albeit not related to community detection, include extending the remap
procedure to other settings such as clustering and mixture model inference.

%% file: appendix-carvalho.tex

\section{Appendix}
\label{app:bsbm}

\subsection{Proof of Theorem~\lowercase{\ref{thm:design}}}
\label{sec:proofdesign}

For the proof we first note that we can split each row $x_{ij}$ in the design
matrix of~\eqref{eq:logitsbm} according to $\gamma$ and $\eta$ entries,
$x_{ij} \doteq [b_{ij} \quad c_{ij}]$, where
\begin{equation}
\label{eq:xbc}
\begin{aligned}
b_{ij,kl} = I[\min(\sigma_i, \sigma_j)=k, \max(\sigma_i, \sigma_j)=l],
&\qquad k,l = 1, \ldots, K, k \le l,\\
c_{ij,v} = I(i=v) + I(j=v),
&\qquad v = 1, \ldots, n,
\end{aligned}
\end{equation}
that is, $b_{ij}$ identifies the pair of communities at the endpoints of
$(i,j)$ for $\gamma$ and $c_{ij}$ marks each node-correction from $\eta$.

\begin{proof}[Proof of (a)]
Let us pick an arbitrary community $k$ and a pair $(i,j)$. There are then
three ways to classify $(i.j)$: (i) it is either outside of community
$k$; (ii) one of its endpoints is in community $k$; or (iii) it is inside
community $k$. If we now define $d_{ij,k} = \sum_{v : \sigma_v = k} c_{ij,v}$
then $(i,j)$ is classified exactly according to $d_{ij,k}$: $d_{ij,k} = 0, 1,$
or $2$ if $(i,j)$ is in cases (i), (ii), or (iii), respectively. Thus, it
follows that
\[
2b_{ij,kk} + \sum_{l\ne k}b_{ij,kl} = \sum_{v : \sigma_v = k} c_{ij,v},
\]
for each $k = 1, \ldots, K$, and so $X$ has $K$ constraints in its columns.
\end{proof}

\begin{proof}[Proof of (b)]
Note that $X$ is full column-ranked if and only if $X^\top X$ is invertible,
so we just need to show that $X^\top X$ is invertible if $N_k \geq 2$ for
$k = 1, \dots, K$.
Let $B=[ b_{ij,12}, \dots, b_{ij,K-1 K}]_{i<j}$ and
$C=[c_{ij,1}, \dots, c_{ij,n}]_{i<j}$. Then $X = [B,C]$ and
\[
X^\top X=  \left[ \begin{array}{cc}
B^\top B & B^\top C \\
C^\top B & C^\top C  \end{array} \right].
\]
Thus, $X^\top X$ is invertible if and only if both $B^\top B$ and the Schur
complement of $C^\top C$,
$\Delta \doteq C^\top[I - B(B^\top B)^{-1}B^\top ]C$ are invertible.
First,
\[
B^\top B = \text{Diag}\Bigg( \sum_{i<j}
I[{\sigma_i = k, \sigma_j=l} \text{ or }{\sigma_i = l, \sigma_j=k}] \Bigg)
=\text{Diag}(N_k N_l),
\]
and so, for this diagonal matrix to be invertible we need $N_k \neq 0$ for $k
= 1, \dots, K$.

As for the Schur complement $\Delta$, we have that
\[
\Delta_{ii} = n-1 - \sum_{k \neq i}\frac{ \sum_{l \neq i} I[\sigma_i\neq
\sigma_k=\sigma_l] }{N_{\sigma_i} N_{\sigma_k}},
\]
and, for $i < j$, 
\[
\Delta_{ij} = 1 - \sum_{k \neq i}\frac{ \sum_{l \neq j} I[\sigma_i=\sigma_j
\neq \sigma_k=\sigma_l \text{ or } \sigma_i=\sigma_l \neq \sigma_k=\sigma_j]
}{N_{\sigma_i} N_{\sigma_k}}.
\]
But if $\sigma_i \neq \sigma_j$,
\[
\Delta_{ij} =1 - \sum_{k \neq i} \frac{\sum_{l \neq j}
I[\sigma_i=\sigma_l \neq \sigma_k=\sigma_j]}{N_{\sigma_i} N_{\sigma_k}} = 0,
\]
and otherwise, if $\sigma_i = \sigma_j$,
\begin{equation}
\label{eq:deltaij}
\Delta_{ij} =1 - \sum_{k \neq i} \frac{\sum_{l \neq i}
I[\sigma_i \neq \sigma_k=\sigma_l]}{N_{\sigma_i} N_{\sigma_k}},
\end{equation}
and so $\Delta_{ii}-\Delta_{ij}=n-2$. Thus, after some row and column
operations, $\Delta$ can be written as a block diagonal matrix where each
block of size $N_k$ has the form:
\[
\left[ \begin{array}{cccc}
p & q &\dots& q \\
q & p  &\dots& q\\
\vdots & & \ddots &\vdots\\
q& q& \dots& p  \end{array} \right]
\]
with $q = \Delta_{ij}$ in~\eqref{eq:deltaij} and $p=n-2+q$. The determinant of
the block diagonal matrix is nonzero if and only if $n \neq 2$ and $N_k \neq
1$. Moreover, the determinant of $X^\top X$ is the same as that of the block
diagonal matrix since one can be obtained from the other through row and
column operations. Thus, the conditions $N_k \neq 0$ from $B^\top B$ and now
$N_k \neq 1$ can be summarized into $N_k \geq 2$.
\end{proof}

\subsection{Remap Algorithm}
\label{sec:remap}
Algorithm~\ref{alg:remap} lists a routine that finds the canonical map $\rho$
based on the canonical order in $\sigma$ as in Equation~(\ref{eq:remap}) and
remaps $\sigma$ in-place.

\begin{algorithm}
  \caption{Remapping labels in $\sigma$ to $\rho(\sigma)$.}
  \begin{algorithmic}
  \label{alg:remap}
    \STATE {\sf assigned} $\leftarrow \{\}$
    \STATE $\rho\leftarrow \{\}$
    \STATE $n \leftarrow 0$ \COMMENT{number of different labels in $\sigma$}
    \FOR[obtain $\rho \doteq$ {\sf ord}$(\sigma)^{-1}$]
      {$i = 1, \ldots, |\sigma|$}
      \IF[first appearance?]{\textbf{not} {\sf assigned}$(\sigma(i))$}
        \STATE {\sf assigned}$(\sigma(i)) \leftarrow$ \textbf{true}
          \COMMENT{mark $\sigma(i)$}
        \STATE $n \leftarrow n + 1$
        \STATE $\rho(\sigma(i)) \leftarrow n$
      \ENDIF
    \ENDFOR
    \FOR[remap $\sigma$]{$i = 1, \ldots, |\sigma|$}
      \STATE $\sigma(i) \leftarrow \rho(\sigma(i))$
    \ENDFOR
    \STATE \textbf{return} $\sigma$
  \end{algorithmic}
\end{algorithm}

\subsection{Proof of Theorem~\lowercase{\ref{thm:centroid}}}
\label{sec:proofcent}

It is sufficient to find the pre-map estimator
\[
\hat{\sigma}^* \doteq \argmin_{\tilde{\sigma} \in \{1,\ldots,K\}^n}
\Exp_{\sigma \given A} \big[ H(\tilde{\sigma}, \rho(\sigma)) \big]
\]
since, by definition, $\hat{\sigma}_C = \rho(\hat{\sigma}^*)$.

Denoting $\Sigma = \{1, \ldots, K\}^n$ and
$\Sigma^* = \Sigma \,/\, \mbox{\sf ord}$, we have that
\[
\begin{split}
\Exp_{\sigma \given A} \big[ H(\tilde{\sigma}, \rho(\sigma)) \big] & =
\sum_{\sigma \in \Sigma} H(\tilde{\sigma}, \rho(\sigma)) \Pr(\sigma \given A) \\
& = \sum_{\sigma \in \Sigma^*} \sum_{\sigma^* : \rho(\sigma^*) = \sigma}
H(\tilde{\sigma}, \sigma) \Pr(\sigma^* \given A). \\
\end{split}
\]

Since $\Pr(\sigma^* \given A) = \Pr(\sigma \given A)$ follows from the lack of
identifiability we further obtain
\[
\Exp_{\sigma \given A} \big[ H(\tilde{\sigma}, \rho(\sigma)) \big] =
\sum_{\sigma \in \Sigma^*} n(\sigma) H(\tilde{\sigma}, \sigma)
\Pr(\sigma \given A),
\]
where $n(\sigma) = |\{\sigma^* : \rho(\sigma^*) = \sigma\}| =
K!/(K-k(\sigma))!$ is the number of assignments that are identified to
$\sigma$ through {\sf ord}, and $k(\sigma)$ is the number of different labels
in $\sigma$. We can then define $\Pr^*(\sigma \given A) \doteq n(\sigma)
\Pr(\sigma \given A)$ as the induced measure in the quotient space $\Sigma^*$
to thus have
\begin{multline*}
\Exp_{\sigma \given A} \big[ H(\tilde{\sigma}, \rho(\sigma)) \big] =
\sum_{\sigma \in \Sigma^*} H(\tilde{\sigma}, \sigma) \Pr^*(\sigma \given A)
= \sum_{\sigma \in \Sigma^*} \sum_{i=1}^n I(\tilde{\sigma}_i \ne \sigma_i)
\Pr^*(\sigma \given A) \\
= n - \sum_{i=1}^n \sum_{\sigma \in \Sigma^*} I(\tilde{\sigma}_i = \sigma_i)
\Pr^*(\sigma \given A)
= n - \sum_{i=1}^n \Pr^*(\sigma_i = \tilde{\sigma}_i \given A).
\end{multline*}

But then
\[
\argmin_{\tilde{\sigma} \in \{1,\ldots,K\}^n}
\Exp_{\sigma \given A} \big[ H(\tilde{\sigma}, \rho(\sigma)) \big]
=
\argmax_{\tilde{\sigma} \in \{1,\ldots,K\}^n}
\sum_{i=1}^n \Pr^*(\sigma_i = \tilde{\sigma}_i \given A)
\]
and so
\[
(\hat{\sigma}^*)_i = \argmax_{k \in \{1, \ldots, K\}}
\Pr^*(\sigma_i = k \given A),
\]
that is, $\hat{\sigma}^*$ is a consensus estimator, as desired.

\subsection{Proof of Theorem~\lowercase{\ref{thm:binder}}}
\label{sec:binder}

To compare $\tilde{\sigma}$ and $\sigma$ let us define
$n_{ij} \doteq \sum_{k, l} I(\sigma_k = i, \tilde{\sigma}_l = j)$,
the number of nodes that belong to community $i$ in $\sigma$ and to community
$j$ in $\tilde{\sigma}$. Then,
$B(\tilde{\sigma}, \sigma) = \sum_{i} \sum_{j < k} (n_{ij} n_{ik}
+ n_{ji} n_{ki})$,
$H(\tilde{\sigma}, \sigma) = \sum_{i \neq j} n_{ij}$, and
$n = \sum_{i,j} n_{ij}$.

For instance, if $K=2$ then $H(\tilde{\sigma}, {\sigma}) = n_{12} + n_{21}$
and 
\[
\begin{split}
B(\tilde{\sigma}, {\sigma}) & = (n_{11}n_{12} + n_{21}n_{22}) + (n_{11}n_{21} + n_{12}n_{22})\\
&= (n_{12}+n_{21})(n_{11}+n_{22})\\
&= H(\tilde{\sigma}, {\sigma})\big(n - H(\tilde{\sigma}, {\sigma})\big).
\end{split}
\]

More generally, for $K > 2$, we have:
\[
\begin{split}
n H(\tilde{\sigma}, {\sigma}) &= \sum_{i \neq j} n_{ij} \sum_{i, j} n_{ij}
= \sum_{i \neq j} n_{ij}
\Bigg( \sum_{i \neq j} n_{ij} + \sum_{i} n_{ii} \Bigg) \\
& = \sum_{i \neq j} n_{ij} \sum_{i \neq j} n_{ij}
+ \sum_{i \neq j} n_{ij} \sum_{i} n_{ii} \\
& = \underbrace{\sum_{i \neq j} n_{ij}^2}_{A}
+ \underbrace{\sum_{\substack{i \neq j, k \neq l \\ k \neq i, j \neq l }}
n_{ij} n_{kl} }_{B} +
2\underbrace{\sum_{\substack{i \neq j, i \neq k \\ j< k }} (n_{ij} n_{ik} +
n_{ji} n_{ki}) }_{C} \\
& \qquad
+ \underbrace{\sum_{\substack{i \neq j, i \neq k \\ j \neq k }} n_{ii}
n_{jk}}_{D}
+ \underbrace{\sum_{i \neq j} (n_{ii} n_{ij} + n_{ii} n_{ji})}_{E}.
\end{split}
\]
Thus, $B(\tilde{\sigma}, \sigma) = C + E$ and, in particular,
\[
\begin{split}
H^2(\tilde{\sigma}, {\sigma}) &=  \Big(  \sum_{ i \neq  j} n_{ij} \Big) \Big(  \sum_{ i \neq j} n_{ij} \Big)\\
& = \sum_{i \neq j} n_{ij}^2 + \sum_{\substack{i \neq j, k \neq l \\ k \neq i, j \neq l }} n_{ij} n_{kl}  +  2\sum_{\substack{i \neq j, i \neq k \\ j< k }} (n_{ij} n_{ik} + n_{ji} n_{ki})  \\
& = A + B + 2C.
\end{split}
\]

The bound 
$B(\tilde{\sigma}, \sigma) \leq H(\tilde{\sigma}, \sigma)
(n - H(\tilde{\sigma}, {\sigma})/2)$ then follows from 
\[
n H(\tilde{\sigma}, {\sigma}) - B(\tilde{\sigma}, {\sigma})
- \frac{1}{2}H^2(\tilde{\sigma}, {\sigma})
= \frac{1}{2}A + \frac{1}{2}B + D \geq 0
\]
since $A,B$ and $D$ are all non-negative.

\begin{figure}
  \includegraphics[width=12cm]{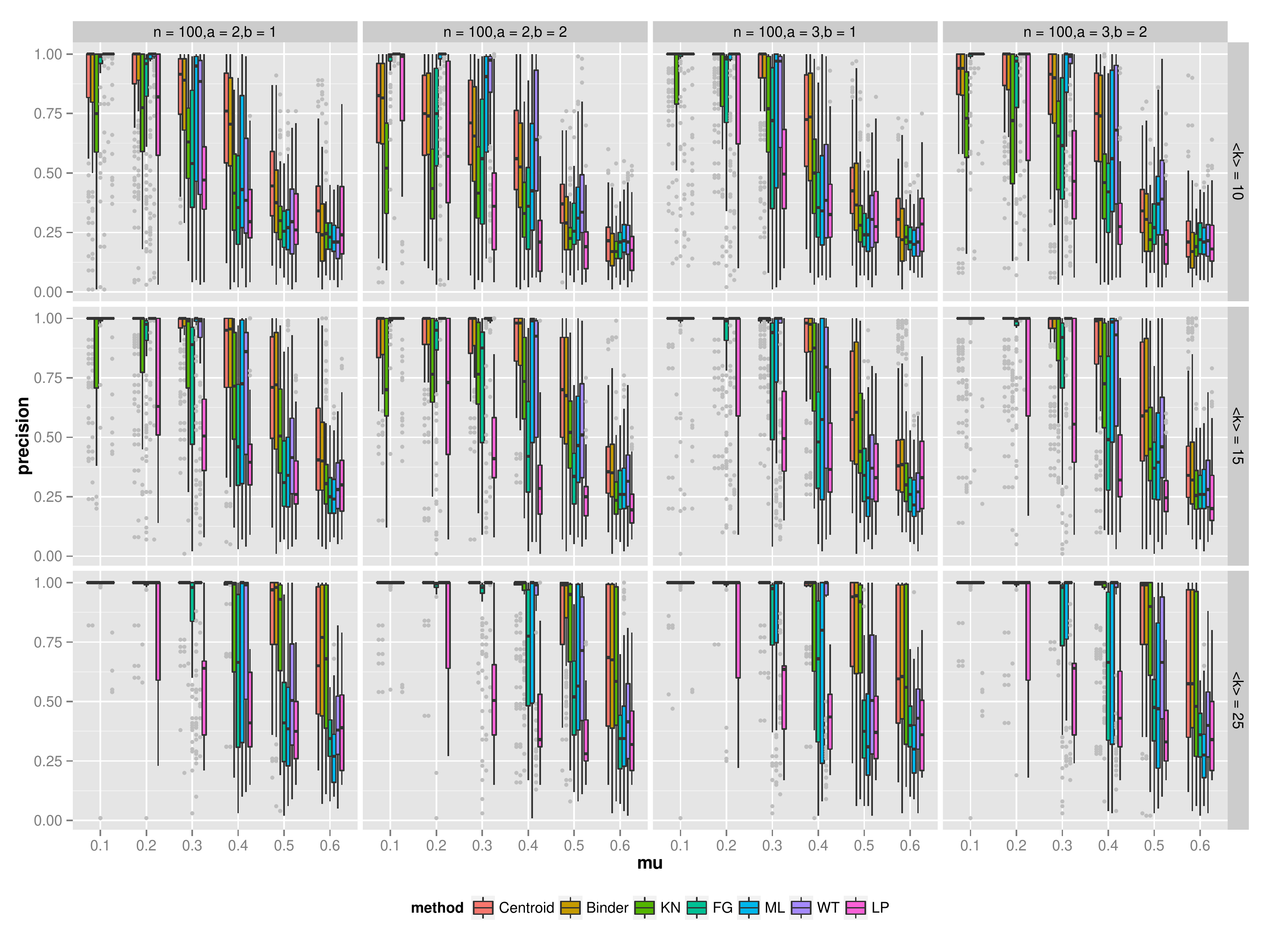}
  \caption{Benchmark networks of $n = 100$ nodes, with different combinations of the exponents $a\in \{2,3\}$, $b\in \{1,2\}$ and the average degree $\langle k \rangle \in \{10, 15, 25\}$ are used. Each boxplot corresponds to the precision of the estimator over 100 graph realizations.
}
 \label{fig:bm100}
\end{figure}

\begin{figure}
  \includegraphics[width=12cm]{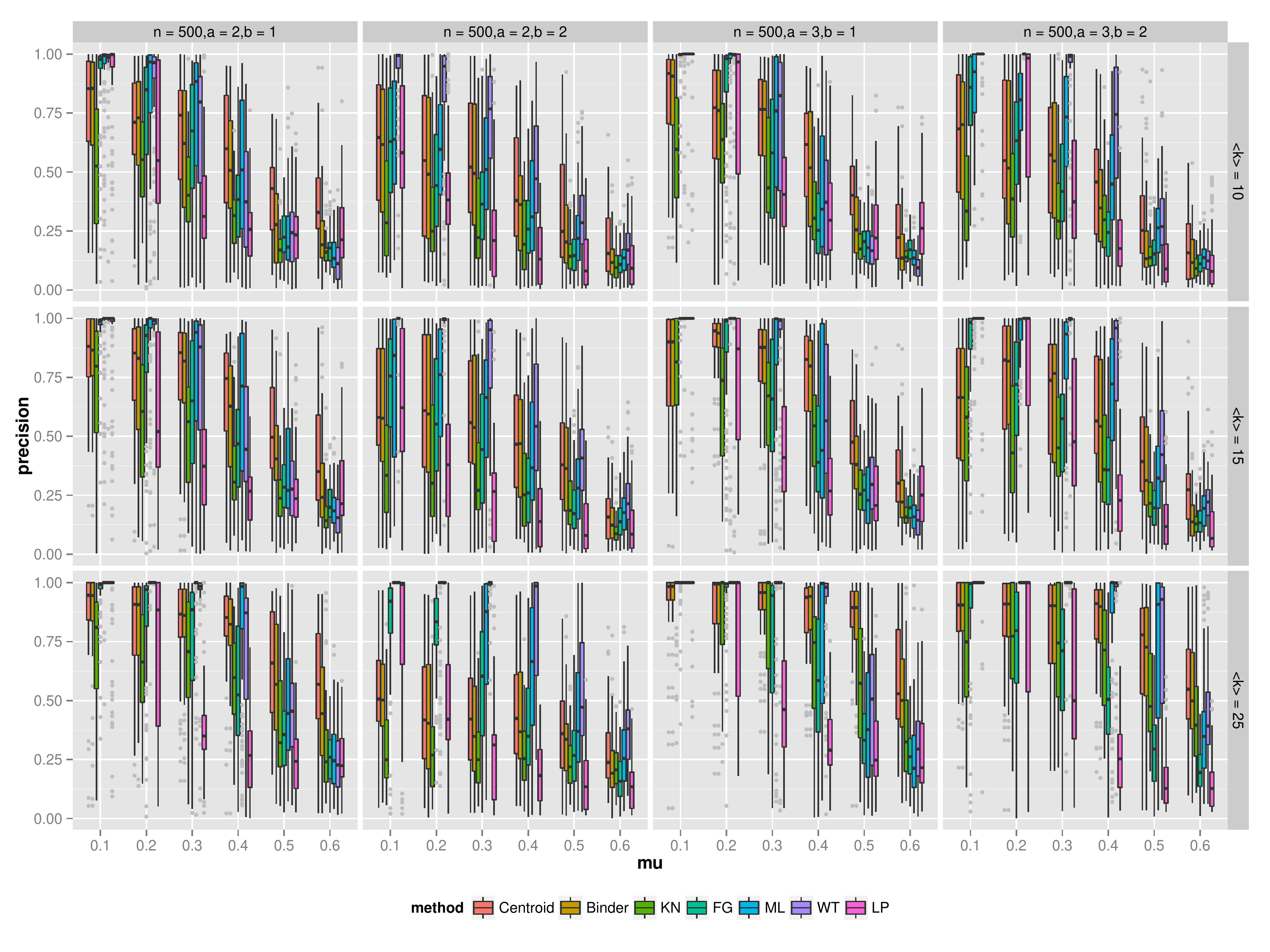}
  \caption{Benchmark networks of $n = 500$ nodes, with different combinations of the exponents $a\in \{2,3\}$, $b\in \{1,2\}$ and the average degree $\langle k \rangle \in \{10, 15, 25\}$ are used. Each boxplot corresponds to the precision of the estimator over 100 graph realizations.
}
 \label{fig:bm500}
\end{figure}